\documentclass[prd,aps,showpacs,nofootinbib,nobibnotes,superscriptaddress]{revtex4}
\usepackage{epsfig,amssymb}
\newcommand{\bea}{\begin{eqnarray}}
\newcommand{\eea}{\end{eqnarray}}
\newcommand{\be}{\begin{equation}}
\newcommand{\ee}{\end{equation}}
\newcommand{\beast}{\begin{eqnarray*}}
\newcommand{\eeast}{\end{eqnarray*}}
\newcommand{\pkt}{\; .}
\newcommand{\kma}{\; ,}
\newcommand{\nn}{\nonumber}

\def\e{{\rm e}}
\newcounter{subequation}[equation]
\makeatletter \expandafter\let\expandafter\reset@font\csname
reset@font\endcsname
  {\endeqnarray\stepcounter{equation}}
\makeatother
\begin{document}

\title{CMB two- and three-point correlation functions from Alfv\'en waves}

\author{Tina Kahniashvili}
\email{tinatin@phys.ksu.edu}
\affiliation{McWilliams Center for Cosmology and Department of Physics,
Carnegie Mellon University, 5000 Forbes Ave, Pittsburgh, PA 15213, USA}
\affiliation{Department of Physics, Laurentian University, Ramsey Lake Road, Sudbury, ON P3E 2C,Canada}
\affiliation{Abastumani Astrophysical Observatory, Ilia State University, 2A Kazbegi Ave, Tbilisi, GE-0160, Georgia}

\author{George Lavrelashvili}
\email{lavrela@itp.unibe.ch}
\affiliation{Department of Theoretical Physics, A.Razmadze Mathematical Institute,
I.Javakhishvili Tbilisi State University, GE-0177 Tbilisi, Georgia}

\date{\today}

\begin{abstract}
We study the cosmic microwave background (CMB) temperature
fluctuations non-gaussianity due to the vector mode perturbations
(Alfv\'en waves) supported by a stochastic cosmological magnetic
field. We present detailed derivations of the statistical properties,
two and three point correlation functions of the vorticity
perturbations and corresponding CMB temperature fluctuations.
\end{abstract}

\maketitle

\section{Introduction}

In the framework of the standard cosmological scenario the cosmic
microwave background (CMB) temperature fluctuations are gaussian and
are fully determined by the CMB temperature fluctuation two-point
correlation functions while higher order odd correlation functions
(for example three-point correlations) are identically zero
(for a review on CMB fluctuations and possible non-gaussianity see
Ref. \cite{review} and references therein). This is a consequence of the
inflationary scenario that predicts the gaussian initial
perturbations, and even at the level of perturbations there is no
violation of rotational symmetry\footnote{The generation of the
vector mode, which involves a preferred direction, is not excluded
during the inflation, but the exponential expansion washes out the
vector (vorticity) perturbations if no supporting external source is
present \cite{mukhanov92}.}. Several observations indicate that the
CMB temperature map could be slightly non-gaussian \cite{WMAP7} and
thus to adequately describe the CMB fluctuations one must study
higher order correlations functions. Furthermore, some modifications
of standard inflationary models lead to a slightly non-gaussian CMB
map \cite{Bernardeau:2002jf}.

A common way to characterize the CMB temperature fluctuations
non-gaussianity is to introduce the $f_{\rm NL}$ parameter, which in
fact determines the relation between the two-point and three-point
correlation functions \cite{Komatsu:2010hc}. The current CMB data
limits $f_{\rm NL} = 32 \pm 21$ \cite{WMAP7}. PLANCK mission will be
able to give us with stronger limits (to improve current limit by an
order or a few). If the nearest future CMB measurements, for example
PLANCK data will confirm that $|f_{\rm NL}|$ is not significantly less
then one, the standard cosmological scenario must be revised
substantially. There are several different ways for such a revision.
For example the inflation could be driven by multiple fields, or the
Universe isotropy has been violated at very early epochs
\cite{review}. Recent studies \cite{Seshadri:2009sy,Caprini:2009vk}
address the magnetic field induced density perturbations as a source
of the CMB temperature fluctuations non-gaussianity. The physical
meaning of this effect is as follows: the temperature anisotropies
caused by the magnetic field are proportional to the magnetic field
energy density parameter, $\Delta T/T \propto B^2/\rho_{\rm cr}$
\cite{Seshadri:2009sy}, where $\rho_{\rm cr} $ is the critical
density today and $B$ is the comoving value of an effective magnetic
field. Accounting that the square of the magnetic field $B^2$ is a
non-linear form \cite{bc05}, the corresponding $\Delta T/T $
fluctuations must be non-gaussian. The limits for the scale
invariant magnetic field amplitude from the CMB non-gaussianity test
is of order of $10^{-9}$ Gauss \cite{Trivedi:2010gi}.

In this paper we investigate the CMB non-gaussianity due to the vector
mode of perturbations induced by a stochastic magnetic magnetic
field, (see Refs. \cite{subramanian98,mkk02,lewis} for details of
the vector magnetized mode). If the magnetic field presence is a
reason for the CMB non-gaussianity, this magnetic field must satisfy
several conditions: (i) the magnetic field must be generated in the
early Universe, prior to recombination; There are different
mechanisms to generate magnetic fields in the early Universe, such
as inflation, phase transitions, see for reviews \cite{origin}; (ii)
the correlation length of the cosmological magnetic field must
satisfy the requirement of causality \cite{durrer}, and thus to have
a field correlated over the horizon scale or even larger this field
should be generated during the inflation and have a scale invariant
spectrum \cite{ratra}; (iii) the amplitude of this magnetic field
should be small enough to preserve the isotropy of the background
Friedman-Lema\^{i}tre-Robertson-Walker (FLRW) metric (so the energy
density of the magnetic field should be the first order of
perturbations), be below the upper bounds (few nGauss) given by
observations \cite{limits}. On the other hand the amplitude of the
magnetic field should be large enough to leave observational traces
on CMB. Recently it was argued that non-observation of blazars in
TeV range by Fermi mission indicates the presence of large scale
correlated intergalactic magnetic field with a lower bound of order
of $10^{-16}$ Gauss \cite{neronov}. The existence of the lower bound
of order of $10^{-16}$ Gauss magnetic field favors the magnetic
field of cosmological origin \cite{dolag}, and thus the magnetic
field amplitude at 1 Mpc is squeezed between $10^{-9}$ and
$10^{-16}$ Gauss.

Vector and tensor modes of magnetized perturbations are much more
complicated then the scalar one. The first paper to address the CMB
bispectrum induced by the vector and tensor mode of perturbations
has been Ref. \cite{Shiraishi:2010sm} where the analytical
expressions were derived. A natural extension of that approach,
namely numerical estimation of the vector and tensor modes induced
non-gaussianity were presented in Refs.
\cite{Shiraishi:2010sm,Shiraishi:2010yk,Shiraishi:2011fi,Shiraishi:2011dh}.
In our study we give detailed derivations of the two and three point
correlation functions of the CMB temperature fluctuations induced by
the magnetized vector mode. In this sense at this stage this article
has a methodological nature.

The outline of the rest of the paper is as follows: In Sec. II we define the
magnetized vector mode of perturbations and compute the Lorentz
force two- (Sec. IIA) and three- (Sec. IIB) point correlation
functions. We explicitly discuss in details the difference of the
vorticity perturbations two-point correlations functions, and show
that the non-gaussianity of the vector field (vorticity) can be seen
already from the two-point correlation function. In Sec. III we
address the CMB temperature fluctuations induced by the magnetized
vector mode perturbations. We present analytical expressions for the
two- (Sec. IIIA) and three- (Sec. IIIB) point correlations of the
CMB temperature anisotropies. We briefly discuss our results and
conclude in Sec. IV. Useful mathematical formulae and  details of
computations are given in Appendix.

\section{Magnetized Perturbations Vector Mode}

To study the dynamics of linear magnetic vector perturbations about
a spatially-flat FLRW homogeneous cosmological spacetime background
(described by the metric tensor ${\bar
g}_{\mu\nu}=a^2\eta_{\mu\nu}$, where
$\eta_{\mu\nu}=\mbox{diag}(-1,1,1,1)$ is the Minkowski metric tensor
and $a(\eta)$ the scale factor) we follow the standard procedure and
decompose the metric tensor into a spatially homogeneous background
part (${\bar g}_{\mu\nu}$) and a perturbation part,
${g}_{\mu\nu}={\bar g}_{\mu\nu}+\delta g_{\mu\nu}$, where $\mu ,\nu
\in (0, 1, 2, 3)$ are spacetime indices. Vector perturbations
$\delta g_{\mu\nu}$ can be described by two three-dimensional
divergence-free vector fields ${\bf A}$ and ${\bf H}$
\cite{mukhanov92}, where
\begin{equation}\label{eq:S-metric-pert}
\delta g_{0i}=\delta g_{i0}=a^2 A_i,\qquad\qquad \delta
g_{ij}=a^2(H_{i,j}+H_{j,i}).
\end{equation}
Here a comma denotes the usual spatial derivative, $i, j \in (1, 2,
3)$ are spatial indices, and ${\bf A}$ and ${\bf H}$ vanish at
spatial infinity. Studying the behavior of these variables under
infinitesimal general coordinate transformations one finds that
${\bf V} = {\bf A}-{\bf \dot H}$ is gauge-invariant (the overdot
represents a derivative with respect to conformal time). ${\bf V}$
is a vector perturbation of the extrinsic curvature \cite{bardeen}.
Exploiting the gauge freedom we choose ${\bf H}$ to be constant in
time. Then the vector metric perturbation may be described in terms
of two divergenceless three-dimensional gauge-invariant vector
fields, the vector potential $\bf{V}$ and a vector representing the
transverse peculiar velocity of the plasma, the vorticity ${\bf
\Omega}={\bf v}-{\bf V}$, where ${\mathbf v}$ is the spatial part of
the four-velocity perturbation of a stationary fluid element
\cite{mkk02}.{\footnote{Given the general coordinate transformation
properties of the velocity field $ \bf v$, two gauge-invariant
quantities can be constructed, the shear ${\bf s} = {\bf v} -{\bf
{\dot H}}$ and the vorticity ${\bf \Omega}= {\bf v } - {\bf A}$
\cite{bardeen}. In the gauge ${\bf \dot H} =0$ (i.e., ${\bf V}={\bf
A}$) we get ${\bf \Omega}={\bf v}-{\bf V}$ \cite{hw97}.}} As we
noted in the Introduction, in the absence of a source the vector
perturbation ${\bf V}$ decays with time (this follows from the
equation for ${\bf \dot V}$, ${\bf \dot V} + 2({\dot a}/{a}) {\bf
V}=0$) and therefore can be ignored.

The residual ionization of the primordial plasma is large enough to
ensure that magnetic field lines are frozen into the plasma.
Neglecting fluid back-reaction onto the magnetic field, the spatial
and temporal dependence of the field separates, ${\mathbf
B}(t,{\mathbf x})={\mathbf B}({\mathbf x})/a^2$ \cite{axel}. Since
the fluid velocity is small the displacement current in Amp\`ere's
law may be neglected; this implies the current $\bf J$ is determined
by the magnetic field via $ {\bf J} ={\bf\nabla} \times {\bf B
}/(4\pi)$.  Accounting for a frozen-in magnetic field lines the
induction law takes the form ${\dot{\bf B}} = {\bf\nabla} \times
({\bf v}\times {\bf B})$. As a result the baryon Euler equation for
$\bf v$ has the Lorentz force ${\mathbf L({\mathbf x})}= - {\mathbf
B}({\bf x})\times\left[\nabla\times {\bf B}({\bf x}) \right]/(4\pi)$
as a source term. The photons are neutral so the photon Euler
equation does not have a Lorentz force source term. The Euler
equations for photons and baryons are \cite{mkk02,lewis}
\begin{eqnarray}\label{eq:V-momentum-baryon}
\dot{{\bf \Omega}}_{\gamma}+\dot{\tau}({\bf v}_{\gamma}-{\bf
v}_{b}) &=& 0,
\label{eq:V-momentum-photon}\\
\dot{{\bf \Omega}}_{b}+\frac{\dot{a}}{a}{\bf \Omega}_{b}-
\frac{\dot{\tau}}{R} ({\bf v}_{\gamma}-{\bf v}_{b}) &=&
\frac{{\bf L}\!^{(V)}\!({\mathbf x})}{a^4(\rho_b+p_b)}~,
\end{eqnarray}
where the subscripts $\gamma$ and $b$ refer to the photon and baryon
fluids, and $\rho$ and $p$ are energy density and pressure. Here
$\dot{\tau}=n_e\sigma_Ta$ is the differential optical depth, $n_e$
is the free electron density, $\sigma_T$  is the Thomson cross
section,
$R=(\rho_b+p_b)/(\rho_\gamma+p_\gamma)\simeq3\rho_b/4\rho_\gamma$ is
the momentum density ratio between baryons and photons, and
$L^{(V)}_i$ is the transverse vector (divergenceless) part of the
Lorentz force. In the tight-coupling limit ${\bf v}_{\gamma}\simeq
{\bf v}_{b}$, so we introduce the photon-baryon fluid divergenceless
vorticity ${\bf \Omega}$ ($={\bf \Omega}_\gamma = {\bf \Omega}_b$)
that satisfies
\begin{equation} \label{vorticity-total}
(1+R)\dot{{\bf \Omega}}+R\frac{\dot{a}}{a}{\bf \Omega}=
\frac{{\bf L}\!^{(V)}\!({\mathbf x})}{a^4(\rho_\gamma+p_\gamma)}.
\end{equation}
The average Lorentz force $\langle{\bf L(x)}\rangle =-\langle{\bf B}
\times [{\bf \nabla} \times {\bf B}]\rangle/(4\pi)$ vanishes, while
the r.m.s. Lorentz force $\langle{\bf L (x) \cdot
L(x)}\rangle^{1/2}$ is non-zero and acts as a source in the vector
perturbation equation.

To proceed one needs to obtain an expression for the Lorentz force
in terms of the magnetic field characteristics. We assume that the
magnetic field is Gaussian and satisfy
\footnote{For a vector field ${\bf F}$ we use
\begin{equation}
   F_j({\mathbf k}) = \int d^3\!{\bf x} \,
   e^{i{\mathbf k}\cdot {\mathbf x}} F_j({\mathbf x}),~~~~~~~~~~~
   F_j({\mathbf x}) = \int {d^3\!{\bf k} \over (2\pi)^3}
   e^{-i{\mathbf k}\cdot {\mathbf x}} F_j({\mathbf k}),\nonumber
\end{equation}
when Fourier transforming between real and wavenumber spaces; we
assume flat spatial hypersurfaces.},
\begin{equation} \label{spectrum}
\langle B_i ({\mathbf k}) B_j^\star ({\mathbf k'})\rangle =
(2\pi)^3 \delta ({\mathbf k}-{\mathbf k'}) P_{ij}({\mathbf{\hat
k}}) M (k)~~~~~{\rm for}~~~~k\leq k_D \kma
\end{equation}
and vanishes for $k>k_D$. Here a star denotes complex conjugation,
and $\delta({\mathbf k}-{\mathbf k'})$ is the usual 3-dimensional
Dirac delta function, and $k_D$ is the magnetic field damping scale
defined through the Alfv\'en waves damping
\cite{subramanian98,jedamzik}.
We approximate power spectrum $M(k)$ by simple power laws.

Following Ref. \cite{mkk02} it can be shown that the Fourier
transform of the Lorentz force $L^{(V)}_i({\bf k})  \equiv k
\Pi_i({\bf k}) $ is related to the Fourier transform of spacial
part of magnetic field energy momentum tensor $\tau^{(B)}_{mj}
(\mathbf{k})$
\be
\tau^{(B)}_{ij} (\mathbf{k}) = \frac{1}{4\pi}
\frac{1}{(2\pi)^3}  \int d^3{\bf q} [B_i (\mathbf{q}) B_j
(\mathbf{k}-\mathbf{q}) -\frac{1}{2}\delta_{ij} B_m (\mathbf{q})
B_m (\mathbf{k}-\mathbf{q})] \kma \label{tau}
\ee
as
\be
\Pi_i(\mathbf{k}) = \frac{1}{2}(P_{ij}({\mathbf{\hat k}})
\hat{k}_m + P_{im}({\mathbf{\hat k}}) \hat{k}_j ) \tau^{(B)}_{mj}
(\mathbf{k})  \kma \label{pi}
\ee
where $P_{ij}({\mathbf{\hat k}}) = \delta_{ij}-\hat{k}_i\hat{k}_j$
is the transverse plane projector with unit wavenumber components
$\hat{k}_i=k_i/k$.

Next we shall solve the Eq. (\ref{vorticity-total}).
It can be solved in two different
regimes, for length scales larger and smaller then comoving Silk
scale $\lambda_S$. These two solutions are \cite{mkk02}: \\
$\star $ $~~$ For the scales $\lambda > \lambda_S$, ($k < k_S)$
\be \label{omega1}
\mathbf{\Omega}(\mathbf{k}, \eta)=
\frac{k \mathbf{\Pi}(\mathbf{k}) \eta }{(1+R)(\rho_{\gamma_0}+p_{\gamma_0})} \kma
\ee
where $\rho_{\gamma_0}$ and  $p_{\gamma_0}$ are photon energy and pressure today.
\\
$\star$ $~~$ For smaller scales with $\lambda < \lambda_S$, $(k >
k_S)$
\be \label{omega2}
\mathbf{\Omega}(\mathbf{k}, \eta)=
\frac{\mathbf{\Pi}(\mathbf{k})}{(kL_\gamma/5)(\rho_{\gamma_0}+p_{\gamma_0})} \kma
\ee
where $L_\gamma$ is the comoving photon mean-free path.

Eqs. (\ref{omega1}) and (\ref{omega2}) define the vorticity through
the magnetic source, $\mathbf{\Pi}({\bf k})$, Eq. (\ref{pi}). As we
can see the vorticity perturbations at scales below the Silk damping
stay constant, while the vorticity perturbations above the Silk
damping are increasing linearly with the time. In order to compute
the magnetized vector mode induced effects we need to compute the
vorticity perturbations two- and three-point correlations.

Below we present our computations. It must be stressed that the
form of the magnetic field two-point correlation function, Eq.
(\ref{spectrum}) presumes the following properties of the field:
(i) transverse, divergence free field, $\mathbf{\nabla} \cdot
{\bf B}=0$, and in the fourier space this is insured by the
projector $P_{ij}({\bf \hat k})$; (ii) isotropy (no preferred
direction) insured by $\delta({\bf k-k'})$ and axis-symmetry of
the field, the two point correlation is symmetric under under $i$
and $j$ indices exchanges;
(iii) The magnetic field is a gaussianly distributed field.


\subsection{Lorentz Force Two-Point Correlation Function}

For the two-point correlation function of the
${\mathbf \Pi }({\bf k})$ magnetic vector source we have
\begin{equation}
\zeta^{(12)}_{i_1 i_2} ({\bf k, k^\prime}) \equiv  \langle
\Pi_{i_1}^\star ({\bf k}) \Pi_{i_2}({\bf k^\prime}) \rangle \pkt
\end{equation}
Using results of Sect. \ref{app_B} one finds for it
\begin{equation} \label{zeta22}
\zeta^{(12)}_{i_1 i_2} ({\bf k, k^\prime}) =
\delta(\mathbf{k}-\mathbf{k^\prime} ) \psi^{(12)}_{i_1 i_2} ({\bf
k})  \kma
\end{equation}
with
\begin{eqnarray}
\psi^{(12)}_{i_1 i_2} ({\bf k} ) & = &  \frac{1}{(8\pi)^2} \int
d^3 {\bf q} M(|{\bf q}|) M(|{\bf k - q}|) \nonumber \\ &&
\left[P_{i_1a}({\bf \hat k}) {\hat k}_b + P_{i_1b}({\bf \hat k})
{\hat k}_a\right] \left[P_{i_2 c}({\bf \hat k}) {\hat k}_d +
P_{i_2d}({\bf \hat k }) {\hat k }_c\right] \left[ P_{ac}({\bf \hat
q}) P_{bd}({\bf{\hat{k-q}}}) + P_{ad} ({\bf \hat q})
P_{bc}({\bf{\hat{k-q}}}) \right]. \label{pi3}
\end{eqnarray}
Note that as usual we assume summation over the repeated indices.
Using $P_{ij}({\bf \hat k})$ projector symmetry properties and
defining $\gamma = {\bf \hat k} \cdot {\bf \hat q} $, $\beta =
{\bf\hat k}\cdot ({\bf {\hat{k-q}}})$, and $\mu = {\bf\hat q}\cdot
({\bf {\hat{k-q}}})$, after long but simple computations we obtain
for the Lorentz force two-point correlation function
\begin{eqnarray} \label{pi4}
\langle L^{(V) \star}_i ({\bf k}) L^{(V)}_j({\bf k^\prime})
\rangle
 & = &  \frac{k^2 P_{i j}({\bf \hat k})}{(8\pi)^2}\delta(\mathbf{k}-\mathbf{k^\prime} )
  \int d^3 {\bf q} M(|{\bf q}|) M(|{\bf
k - q}|)(2-\beta^2 - \gamma^2) + \nonumber\\
& + & \frac{k_i k_j}{(8\pi)^2}\delta(\mathbf{k}-\mathbf{k^\prime}
) \int d^3 {\bf q} M(|{\bf q}|) M(|{\bf k - q}|) \left[2\gamma^2
\beta^2 - \gamma^2 (1-\beta^2) -
\beta^2(1-\gamma^2) \right] \nonumber \\
& - & \frac{k^2}{(8\pi)^2}\delta(\mathbf{k}-\mathbf{k^\prime} )
  \int d^3 {\bf q} M(|{\bf q}|) M(|{\bf
k - q}|) \left\{ {\hat q}_i {\hat q}_j (1-\beta^2) +  ({\bf
{\hat{k-q}}})_i ({\bf {\hat{k-q}}})_j (1-\gamma^2) \right.
\nonumber \\
& - & \gamma ({\hat k}_i {\hat q}_j + {\hat k}_j {\hat q}_i)
(1-2\beta^2)
 - \beta \left[ {\hat k}_i ({\bf {\hat{k-q}}})_j  + {\hat k}_j ({\bf {\hat{k-q}}})_i \right] (1-2\gamma^2)
\nonumber \\ & - &
 \gamma \beta \left[ {\hat q}_i ({\bf {\hat{k-q}}})_j + {\hat q}_j ({\bf
 {\hat{k-q}}})_i \right] \left. \right\} .
\end{eqnarray}
Note that the form of the Lorentz force two-point correlation
function,  Eq. (\ref{pi4}), has several symmetries: (i) symmetry
under $i$ and $j$ index exchange; (ii) the symmetry under ${\bf \hat
q}$ and ${\bf\hat{k-q}}$ exchange (i.e. $\gamma$ and $\beta$ angles
exchange symmetry). It must be underlined that the trace of
Eq.~(\ref{pi4}) is given by
\begin{equation}
\langle L^{(V) \star}_i ({\bf k}) L^{(V)}_i({\bf k^\prime})\rangle
 =   \frac{k^2}{2 (4\pi)^2}\delta(\mathbf{k}-\mathbf{k^\prime} )
\int d^3 {\bf q} M(|{\bf q}|) M(|{\bf k - q}|)
(1+\mu\gamma\beta - 2\beta^2\gamma^2) \kma
\end{equation}
which totally agrees with the result of Refs.
\cite{mkk02,lewis,bc05}. Further simplification of Eq.~(\ref{pi4})
gives following result for the Lorentz force two-point correlation
function:
\begin{equation}\label{piX}
\langle L^{(V) \star}_i ({\bf k}) L^{(V)}_j({\bf k^\prime})\rangle
 =   \frac{k^2}{(8\pi)^2} P_{i j}({\bf \hat k}) \delta(\mathbf{k}-\mathbf{k^\prime} )
\int d^3 {\bf q} M(|{\bf q}|) M(|{\bf k - q}|)
(1+\mu\gamma\beta - 2\beta^2\gamma^2) \pkt
\end{equation}
The ensemble averaging procedure insures that rotational isotropy is
preserved. Note that without ensemble averaging {\it in one
realization} the isotropy of the two point correlation can be
violated and only the averaging leads to cancelation of anisotropic
component and restoration of isotropy.

\subsection{Lorentz Force Three-Point Correlation Function}

The three-point correlation function of the vector magnetic source is
given\footnote{It is appropriate to define the three point correlation function
as an average of three ${\Pi}_{i}$'s without the complex conjugation.} by
\begin{equation} \label{zeta1}
\zeta^{(123)}_{i_1 i_2 i_3} ({\bf k_1, k_2, k_3})= \langle
{\Pi}_{i_1}(\mathbf{k_1}) {\Pi}_{i_2}(\mathbf{k_2})
{\Pi}_{i_3}(\mathbf{k_3}) \rangle \pkt
\end{equation}
Using results of Sec. \ref{app_B} one finds
\begin{equation} \label{zeta2}
\zeta^{(123)}_{i_1 i_2 i_3} ({\bf k_1, k_2, k_3})=
\delta(\mathbf{k_1}+\mathbf{k_2}+\mathbf{k_3} ) \psi^{(123)}_{i_1
i_2 i_3} ({\bf k_1, k_2})\kma
\end{equation}
with
\bea \label{psi} \psi^{(123)}_{i_1 i_2 i_3}({\bf k_1, k_2})&=&\frac{1}{8 \pi^3}
\int d^3 \mathbf{q}
M(|\mathbf{q}|) M(|\mathbf{k}_1-\mathbf{q}|) M(|\mathbf{k}_2+\mathbf{q}|) \nn \\
&& P_{i_1 j_1} (\mathbf{\hat k}_1) P_{j_1 s_3} (\hat{\mathbf{k}_1
- \mathbf{q}}) ({\bf \hat{k_1+k_2}})_{ {s_3}} P_{i_2 j_2} (\mathbf{\hat k}_2)
P_{j_2 j_3} (\hat{\mathbf{k}_2 + \mathbf{q}}) P_{ i_3 j_3}({\bf \hat{k_1+k_2}})
\hat{k}_{1 {s_1}} P_{s_1 s_2} (\mathbf{\hat
q}) \hat{k}_{2 {s_2}} \pkt \eea
The Lorentz force three point correlation function is given as
\bea \label{psi2}\langle L^{(V)}_{i_1}(\mathbf{k_1})
L^{(V)}_{i_2}(\mathbf{k_2}) L^{(V)}_{i_3}(\mathbf{k_3}) \rangle
&=&\frac{1}{8 \pi^3}
\delta(\mathbf{k_1}+\mathbf{k_2}+\mathbf{k_3} ) \int d^3
\mathbf{q}
M(|\mathbf{q}|) M(|\mathbf{k}_1-\mathbf{q}|) M(|\mathbf{k}_2+\mathbf{q}|) \nn \\
&& P_{i_1 j_1} (\mathbf{\hat k}_1) P_{i_2 j_2} (\mathbf{\hat k}_2)
P_{ i_3 j_3} (\mathbf{\hat{{\bf k_1} + {\bf k_2}}}) P_{j_1 s_3}
(\hat{\bf{k_1-q}}) ({\bf{{k_1 +k_2}}})_{s_3}  P_{j_2 j_3}
(\hat{\mathbf{k}_2 + \mathbf{q}}) \nonumber \\ & & \left[({\bf k_1}
\times {\bf\hat q} )({\bf k_2 } \times {\bf\hat q} ) \right] \pkt
\eea We see that the r.h.s. of Eq. (\ref{psi2}) is symmetric under
exchange ${\bf k_1}$ and ${\bf k_2}$. It is obvious that this
symmetry is reflected in the CMB temperature three point correlation
function, see below.

\section{Temperature fluctuations from the magnetized vector mode}

Vector perturbations induce CMB temperature anisotropies via the
Doppler and integrated Sachs-Wolfe effects \cite{dky98,klr08},
\begin{equation}
\frac{\Delta T}{T}({\bf x_0, n}, \eta_0) = -{\mathbf
v}\cdot{{\mathbf n}}|^{\eta_0}_{\eta_{\text{dec}}} +
\int^{\eta_0}_{\eta_{\text{dec}}} d\eta\,\dot{{\mathbf
V}}\cdot{{\mathbf n}}, \label{eq:V-CMB-1}
\end{equation}
where $\mathbf{n}$ is the unit vector in the light propagation
direction, $\eta_{\text{dec}}$ is the conformal time at decoupling.
${\bf x_0}$ denotes the observer position, $ \mathbf{x_{\rm dec}}=
{\bf x_0} + \mathbf{n} (\eta_0-\eta_{\rm dec})$. Due to the
spherical symmetry the temperature fluctuations are decomposed using
the spherical harmonics as,
\begin{equation}
\frac{\Delta T}{T} (\mathbf{x_0, n, \eta_0}) = \sum_{l=0}^\infty \sum_{m=-l}^l a_{lm} ({\bf x_0},
\eta_0)\cdot Y_{lm}({\bf n}) \pkt
\end{equation}
Accounting for the spherical harmonics property, see Chapter 5 of
Ref.~\cite{varshalovich89}, we obtain $a_{lm}({\bf x_0}, \eta_0) =
\int d\Omega_{\bf n} Y^\star_{lm}({\bf n}) \Delta T/T({\bf x_0, n},
\eta_0)$. The decaying nature of the vector potential ${\mathbf V}$
implies that most of its contribution toward the integrated
Sachs-Wolfe term comes from near $\eta_{\text{dec}}$. Neglecting a
possible dipole contribution due to ${\mathbf v}$ today, from
Eq.~(\ref{eq:V-CMB-1}) we obtain,
\begin{equation} \label{eq:V-CMB-2}
\frac{\Delta T}{T}({{\mathbf n}}) \simeq {\mathbf
v}(\eta_{\text{dec}}) \cdot{{\mathbf n}}-{\mathbf
V}(\eta_{\text{dec}})\cdot{{\mathbf n}} =  {\bf \Omega}(\eta_{\rm
dec}) \cdot {\bf n} \pkt
\end{equation}
Here we placed observer at the origin
$\mathbf{x_0}=0$ and we skip $\eta_0$ denoting
$
{\Delta T}/{T}({\bf n}) \equiv {\Delta
T}/{T}({\bf x_0} =0, {\bf n}, \eta=\eta_0)$ and $a_{lm} \equiv a_{lm}({\bf x_0}=0, \eta_0)$.
Since $ \mathbf{x_{\rm dec}}= \mathbf{n} (\eta_0-\eta_{\rm dec})$
Fourier transform of Eq.~(\ref{eq:V-CMB-2}) results in
\begin{equation} \label{DTdef}
\frac{\Delta T}{T}({\bf k, n})=
{\bf \Omega }  ({\bf k}, \eta_{\rm dec}) \cdot {\bf n})
e^{i({\bf k} \cdot {\bf n})\Delta \eta} \kma
\end{equation}
where ${\bf \Omega}({\bf k}, \eta_{\rm dec})$ is the Fourier
amplitude of vorticity perturbations at $\eta= \eta_{\rm dec}$, wave
vector ${\bf k}= k {\bf{\hat k}}$ labels the resulting Fourier mode
after transforming from the coordinate representation ${\bf x}$ to
the momentum representation by using $\e^{i {\bf k x}}$, and $\Delta
\eta =\eta_0-\eta_{\rm dec} \approx \eta_0$ is the conformal time
from decoupling until today.

For the multipole coefficient Fourier transform we
obtain, \be a_{l m} ({\mathbf k}) = \int d{{\mathbf n}}
(\mathbf{\Omega}(\mathbf{k}, \eta_{\rm dec}) \cdot \mathbf{n})
e^{i\mathbf{k\cdot n}\eta_0} Y^\star_{lm} ({\hat{\mathbf n}})
\label{alm}\kma \ee where we use $ a_{lm} ({\bf k}) \equiv
a_{lm}(\mathbf{k}, \eta=\eta_0)$.

\subsection{Two-point correlation function}

The two-point correlation of the temperature fluctuations is given by
\begin{equation}
{\mathcal C} (\mathbf{n_1}, \mathbf{n_2})\equiv
\langle \frac{\Delta T}{T}(\mathbf{n_1}) \frac{\Delta T}{T} (\mathbf{n_2}) \rangle
= \sum_{l_1 m_1} \sum_{l_2 m_2} \langle a_{l_1 m_1}^\star a_{l_2 m_2}  \rangle
Y_{l_1 m_1}^\star(\mathbf{n_1}) Y_{l_2 m_2}(\mathbf{n_2})\kma
\end{equation}
here $\sum_{l m} \equiv \sum_{l=0}^{\infty} \sum_{m=-l}^l$. In the
case when the spatial isotropy and rotational invariance is
preserved the multipoles with different $l $ and $m$ do not
correlate and $\langle a_{l_1 m_1}^\star a_{l_2 m_2} \rangle =
C_{l_1} \delta_{l_1 l_2} \delta_{m_1 m_2}$. It is obvious that
${\mathcal C} (\mathbf{n_1}, \mathbf{n_2})$ is the function of the
angle between ${\bf n_1}$ and ${\bf n_2}$ vectors, and we can easy
get
\begin{equation}
{\mathcal C} (\mathbf{n_1}, \mathbf{n_2}) = \sum_{l m} \frac{2l+1}{4\pi} C_l P_l({\bf n_1 \cdot n_2}).
\end{equation}
The sound waves (density perturbations) sourced by the magnetic
field keep the rotational invariance unchanged and thus the
multipole correlation matrix has a diagonal form, and the two-point
correlation function depends on one angle, ${\bf n_1 \cdot n_2}$.
The same time the temperature fluctuations are non-gaussian
\cite{Seshadri:2009sy,Caprini:2009vk} due to the non-linearity of
the magnetic field energy density with respect to the magnetic
field. Note, that the magnetic field which has a non-gaussian energy
momentum tensor \cite{bc05} can be itself field with Gaussian
distribution.

As soon as the vector mode is present in the Universe there is an
additional direction inserted through the vorticity field ${\bf
\Omega}(\eta_{\rm dec})$. Before ensemble averaging procedure the
two-point correlation is not rotationally invariant and the cross
correlation function between the multipoles has off-diagonal
components \cite{dky98,klr08}. The measurement of these off-diagonal
terms might serve as a tool to constraint the primordial homogeneous
magnetic field.

We define the multipole two point correlation function as usual:
\be
D^{m_1 m_2}_{l_1 l_2} \equiv \langle a_{l_1 m_1}^\star a_{l_2
m_2} \rangle  = \int \frac{d^3 {\mathbf k_1}}{(2\pi)^3} \frac{d^3
{\mathbf k_2}}{(2\pi)^3}  \langle a_{l_1 m_1}^\star ({\mathbf k_1})
a_{l_2 m_2} ({\mathbf k_2})  \rangle \kma
\ee
and the simple calculation leads to
\bea \label{bispectrum11}
\langle a_{l_1 m_1}^\star a_{l_2 m_2}  \rangle =\int \frac{d^3
{\mathbf k_1}}{(2\pi)^3} \frac{d^3 {\mathbf k_2}}{(2\pi)^3}
d{{\mathbf n_1}} d{{\mathbf n_2}} e^{i ( -\mathbf{k_1 \cdot
n_1}+\mathbf{k_2 \cdot n_2}) \eta_0} Y_{l_1 m_1}^\star ({{\mathbf
n_1}}) Y_{l_2 m_2} ({{\mathbf n_2}}) \langle [{\bf n_1}
\cdot \mathbf{\Omega^\star }(\mathbf{k_1}, \eta_{\rm dec}) ][{\bf
n_2} \cdot \mathbf{\Omega}(\mathbf{k_2}, \eta_{\rm dec})] \rangle
\pkt
\eea
We also decompose the vorticity perturbation plane wave over the
vector spherical harmonics as, $\mathbf{\Omega}({\bf k}) e^{i{\bf k
\cdot n} \eta_0}= \sum_{\lambda, l,m} A^{(\lambda )}_{lm} {\bf
Y}_{lm}^{(\lambda)}({\bf n})$, with $\lambda =-1, 0, 1$ and use
${\mathbf{\nabla} \cdot \mathbf{\Omega} =0}$ condition, leading to
$\mathbf{\Omega}({\bf k}) \cdot {\bf Y}_{lm}^{(-1)}({\bf \hat
k})=0$, see also Appendix of Ref. \cite{klr08}.  We use also the
definition ${\mathbf n} \cdot Y^\star_{l m} ({{\mathbf n}}) =
{\mathbf Y}^{(-1) \star}_{l m} ({{\mathbf n}}) $. Then we have
\begin{equation}
D^{m_1 m_2}_{l_1 l_2}  = \int \frac{d^3 {\mathbf k_1}}{(2\pi)^3}
\frac{d^3 {\mathbf k_2}}{(2\pi)^3} d{{\mathbf n_1}} d{{\mathbf
n_2}} Y^\star_{l_1 m_1} ({{\mathbf n_1}}) Y_{l_2 m_2} ({{\mathbf
n_2}}) \sum_{t_1 r_1} \sum _{t_2 r_2} Y^\star_{t_1 r_1} ({{\mathbf
n_1}}) Y_{t_2 r_2}({{\mathbf n_2}}) \langle A^{(-1)
\star}_{t_1 r_1} A^{(-1)}_{t_2 r_2}  \rangle \kma \label{int1}
\end{equation}
with
\begin{equation}
A^{(-1)}_{lm}  = 4\pi i^{l-1} \sqrt{l(l+1)}
\frac{j_l(k\eta_0)}{k\eta_0} (\mathbf{\Omega}({\bf k}) \cdot {\bf
Y}^{(+1)}_{lm} ({\bf \hat k})) \pkt
\end{equation}
The integration over ${\bf n}$ in Eq.~(\ref{int1}) gives us
$\delta_{l_1 t_1}$, $\delta_{l_2 t_2}$, $\delta_{m_1 r_1}$,
$\delta_{m_2 r_2}$, and the sums are eliminated and we get
\begin{eqnarray}
D^{m_1 m_2}_{l_1 l_2}  &=& \frac{ (-1)^{l_1-l_2} ~i^{l_1+l_2-2}
\sqrt{l_1(l_1+1) l_2(l_2+1)}}{(2\pi^2)^2} \int {d^3 {\mathbf
k_1}} ~{d^3 {\mathbf k_2}} \frac{j_{l_1}(k_1 \eta_0)
j_{l_2}(k_2\eta_0)}{k_1k_2 \eta_0^2 }\nonumber\\ && \left[ {\bf
Y}^{(+1)\star}_{l_1m_1} ({\bf \hat k_1}) \right]_{i_1} \left[{\bf
Y}^{(+1)}_{l_2m_2} ({\bf \hat k_2})\right]_{i_2} \langle
{\Omega}_{i_1}^\star({\bf k_1}) {\Omega}_{i_2}({\bf k_2}) \rangle \pkt
\label{int2}
\end{eqnarray}
Next we have to consider separately case of large scale approximation
Eq.~(\ref{omega1}) and case of small scale approximation Eq.~(\ref{omega2}).
Using  Eq. (\ref{piX}) we obtain:
\begin{eqnarray} \label{inte3}
D^{m_1 m_2}_{l_1 l_2}   &=& \frac{ (-1)^{l_1-l_2} ~i^{l_1+l_2-2}
\sqrt{l_1(l_1+1) l_2(l_2+1)}}{(2\pi)^3 [2(1+R_{\rm dec})(\rho_{\gamma_0}
+ p_{\gamma_0})]^2 }  \left(\frac{\eta_{\rm dec}}{\eta_0} \right)^2
\int {d^3 {\mathbf k}} ~{d^3 {\mathbf q}} M(|{\bf q}|)
M(|{\bf k-q}|) j_{l_1}(k \eta_0) j_{l_2}(k\eta_0)\nonumber\\ &&
  (1+\mu\gamma\beta-2\gamma^2\beta^2) \left( {\bf Y}^{(+1) \star}_{l_1 m_1}
({\bf\hat k}) \cdot  {\bf Y}^{(+1)}_{l_2
m_2}({\bf\hat k}) \right)
\end{eqnarray}
for the scales larger then the Silk damping scale.
For the scale smaller than the Silk damping scale we have
\begin{eqnarray}\label{inte3s}
D^{m_1 m_2}_{l_1 l_2}   &=& \frac{ (-1)^{l_1-l_2} ~i^{l_1+l_2-2}
\sqrt{l_1(l_1+1) l_2(l_2+1)}}{(2\pi)^3 [2(L_\gamma /5)(\rho_{\gamma_0}
+ p_{\gamma_0})]^2 } \int {d^3 {\mathbf k}} ~{d^3 {\mathbf q}}
M(|{\bf q}|) M(|{\bf k-q}|) \frac{j_{l_1}(k \eta_0)
j_{l_2}(k\eta_0)}{(k \eta_0)^2 }\nonumber\\
&&
  (1+\mu\gamma\beta-2\gamma^2\beta^2) \left( {\bf Y}^{(+1) \star}_{l_1 m_1}
({\bf\hat k}) \cdot  {\bf Y}^{(+1)}_{l_2
m_2}({\bf\hat k}) \right) \pkt
\end{eqnarray}
Recall that we assume that the magnetic field power spectrum is
given by a simple power law, $M(|{\bf q}|) \propto q^n$. To proceed
we have to evaluate the following integrals over angular variables
\begin{eqnarray}
{\mathcal I}_{l_1, l_2}^{m_1, m_2} = \int d\Omega_{\bf {\hat k}} \int d\Omega_{\bf{\hat q}}
(1+\mu\beta\gamma-2\gamma^2\beta^2)(k^2+q^2-2kq\gamma)^{n/2}
\left( {\bf Y}^{(+1) \star}_{l_1 m_1}
({\bf\hat k}) \cdot  {\bf Y}^{(+1)}_{l_2
m_2}({\bf\hat k}) \right)~.
\label{angular-int}
\end{eqnarray}
It is easy to see that integrals over $d\Omega_{\bf{\hat k}}$ and
$d\Omega_{\bf{\hat q}}$ separate. First we evaluate the integral
over $d\Omega_{\bf{\hat q}}$. For a particular ${\bf\hat {k}}$,
choose the polar axis for the $d\Omega_{\bf{\hat q}}$ integral in
the ${\bf z}$ direction. Since $\gamma=\cos \theta_{\bf {\hat q}}$
the integrand does not depend on the azimuthal angle $\phi_{\bf
{\hat q}}$ and the integration over $\phi_{\bf {\hat q}}$ simply
gives $2\pi$. The integration over $d \cos \theta_{\bf {\hat q}}$ is
simple to be evaluated and is given as
$
 \int_{-1}^{1} d\gamma (1-\gamma^2) \left ( 1-\frac{2q\gamma(k+q\gamma)}{k^2+q^2-2kq\gamma} \right)
 \left(k^2+q^2-2kq\gamma\right)^{n/2},
$
with $q=|{\bf q}|$. Thus
\begin{eqnarray}
{\mathcal I}_{l_1, l_2}^{m_1, m_2} = 2\pi \delta_{l_1 l_2} \delta_{m_1 m_2} \int_{-1}^{1} d\gamma (1-\gamma^2)
\left ( 1-\frac{2q\gamma(k+q\gamma)}{k^2+q^2-2kq\gamma} \right) \left(k^2+q^2-2kq\gamma\right)^{n/2}
\label{angular-int2}
\end{eqnarray}
and finally only the diagonal cross correlations are present. Note
that, the rotational symmetry (absence of off-diagonal terms in
$D_{l_1 l_2}^{m_1 m_2}$ ) is due to the ensemble averaging.

\subsection{Bispectrum definition and calculation}

The standard approach to study the CMB non-gaussianity consists on
the CMB temperature fluctuations investigation. Below we present the
self-consistent way to describe the bispectrum. As usual the
three-point correlation function of CMB temperature anisotropy is
defined as
\begin{equation}
\xi (\mathbf{n_1}, \mathbf{n_2}, \mathbf{n_3})\equiv
\langle \frac{\Delta T}{T}(\mathbf{n_1}) \frac{\Delta T}{T} (\mathbf{n_2})
\frac{\Delta T}{T} (\mathbf{n_3}) \rangle
= \sum_{l_i m_i} \langle a_{l_1 m_1} a_{l_2 m_2} a_{l_3 m_3} \rangle
Y_{l_1 m_1}(\mathbf{n_1}) Y_{l_2 m_2}(\mathbf{n_2}) Y_{l_3 m_3}(\mathbf{n_3}) \kma i=1,2,3 \kma
\end{equation}
In order to proceed we need to calculate the following form
\be
B^{m_1 m_2 m_3}_{l_1 l_2 l_3} \equiv
\langle a_{l_1 m_1} a_{l_2 m_2} a_{l_3 m_3} \rangle  = \int
\frac{d^3 {\mathbf k_1}}{(2\pi)^3} \frac{d^3 {\mathbf k_2}}{(2\pi)^3} \frac{d^3 {\mathbf k_3}}{(2\pi)^3}
\langle a_{l_1 m_1} ({\mathbf k_1}) a_{l_2 m_2} ({\mathbf k_2}) a_{l_3 m_3} ({\mathbf k_3}) \rangle
\pkt
\ee
Using expression for $a_{lm}$ given by Eq.~(\ref{alm}) we get
\bea \label{bispectrum1}
\langle a_{l_1 m_1} a_{l_2 m_2} a_{l_3 m_3} \rangle =\int
\frac{d^3 {\mathbf k_1}}{(2\pi)^3} \frac{d^3 {\mathbf k_2}}{(2\pi)^3} \frac{d^3 {\mathbf k_3}}{(2\pi)^3}
d{{\mathbf n_1}} d{{\mathbf n_2}} d{{\mathbf n_3}}
e^{i ( \mathbf{k_1 \cdot n_1}+\mathbf{k_2 \cdot n_2}+\mathbf{k_3 \cdot n_3} )\Delta \eta} \nn \\
Y^\star_{l_1 m_1} ({{\mathbf n_1}})
Y^\star_{l_2 m_2} ({{\mathbf n_2}})
Y^\star_{l_3 m_3} ({{\mathbf n_3}})
n_{1 {i_1}} n_{2 {i_2}} n_{3 {i_3}} \langle
{\Omega}_{i_1}(\mathbf{k_1}, \eta_{\rm dec})
{\Omega}_{i_2}(\mathbf{k_2}, \eta_{\rm dec})
{\Omega}_{i_3}(\mathbf{k_3}, \eta_{\rm dec}) \rangle \pkt
\eea
First we integrate over $d{\mathbf n}_i$ by using Eq.~(117) on
p.227 \cite{varshalovich89}. Proceeding in the way given in Sec.
IIIA we arrive at
\bea
&&B^{m_1 m_2 m_3}_{l_1 l_2 l_3} =
\frac{i^{l_1+l_2+l_3-3}}{(2\pi^2)^3 } \sqrt{l_1(l_1+1) l_2(l_2+1)
l_3(l_3+1)} \int d^3{\bf k_1} d^3{\bf k_1} d^3{\bf k_3}
\frac{j_{l_1}(k_1 \eta_0) j_{l_2}(k_2 \eta_0) j_{l_3}(k_3 \eta_0)}{k_1k_2k_3 {\eta_0}^3} \nn \\
&& {{\mathbf Y}^{(+1)}_{l_1 m_1}}^\star (\hat{\mathbf
k}_1)|_{i_1} {{\mathbf Y}^{(+1)}_{l_2 m_2}}^\star (\hat{\mathbf
k}_2)|_{i_2} {{\mathbf Y}^{(+1)}_{l_3 m_3}}^\star (\hat{\mathbf
k}_3)|_{i_3} \langle {\Omega}_{i_1}(\mathbf{k_1}, \eta_{\rm dec})
{\Omega}_{i_2}(\mathbf{k_2}, \eta_{\rm dec})
{\Omega}_{i_3}(\mathbf{k_3}, \eta_{\rm dec}) \rangle \pkt
\eea
Next we have to consider separately case of large scale approximation Eq.~(\ref{omega1})
and case of small scale approximation  Eq.~(\ref{omega2}).
Using Eqs.~(\ref{zeta1}, \ref{zeta2}, \ref{psi}), relation Eq.~(\ref{abcd})
and the representation for the Dirac delta function Eq.~(\ref{delta})
after some computations  for the large scale approximation ($L>L_S$) we obtain
\bea \label{bispectrum5}
B^{m_1 m_2 m_3}_{l_1 l_2 l_3} &=&
\frac{i^{l_1+l_2+l_3-3}\sqrt{l_1(l_1+1) l_2(l_2+1)
l_3(l_3+1)}}{(2\pi^3)^3 [(1+R_{\rm dec})(\rho_{\gamma_0} +
p_{\gamma_0})]^3 } \left(\frac{\eta_{\rm dec}}{\eta_0} \right)^3
 \int k_1^2 dk_1 k_2^2 dk_2  k_3^2 dk_3
j_{l_1}(k_1 \eta_0) j_{l_2}(k_2 \eta_0) j_{l_3}(k_3 \eta_0) \nn \\
&&\int d\Omega_{\hat{\mathbf k}_1} d\Omega_{\hat{\mathbf k}_2} d\Omega_{\hat{\mathbf k}_3}
\int d\Omega_{\hat{\mathbf q}} \int q^2 dq
M(|{\mathbf q}|) M(|{\mathbf k}_1-{\mathbf q}|) M(|{\mathbf k}_2+{\mathbf q}|) \nn \\
&&\sum_{t_1 r_1} \sum_{t_2 r_2} \sum_{t_3 r_3} i^{t_1+t_2+t_3} \int x^2 dx
j_{t_1}(k_1 x) j_{t_2}(k_2 x)  j_{t_3}(k_3 x)
 \ {G^{t_1 t_2 t_3}_{r_1 r_2 r_3}} \nn \\
&&\left[ {\mathbf Y}^{(-1)}_{t_1 r_1} (\hat{\mathbf k}_1) \times {\hat{\mathbf q}}\right]_a
\left[{{\mathbf Y}^{(1)}_{l_1 m_1}}^\star (\hat{\mathbf k}_1)\times(\hat{{\mathbf k}_1-\mathbf q})\right]_b \nn \\
&&\left[{\mathbf Y}^{(-1)}_{t_2 r_2} (\hat{\mathbf k}_2) \times \hat{\mathbf q}\right]_a
\left[{{\mathbf Y}^{(1)}_{l_2 m_2}}^\star (\hat{\mathbf k}_2) \times (\hat{{\mathbf k}_2+\mathbf q})\right]_c \nn \\
&&\left[{\mathbf Y}^{(-1)}_{t_3 r_3} (\hat{\mathbf k}_3) \times (\hat{{\mathbf k}_1 -\mathbf q})\right]_b
\left[{{\mathbf Y}^{(1)}_{l_3 m_3}}^\star (\hat{\mathbf k}_3) \times (\hat{{\mathbf k}_2+\mathbf q})\right]_c
\kma
\eea
where $G^{t_1 t_2 t_3}_{r_1 r_2 r_3}$ is the usual Gaunt integral,
\be
G^{t_1 t_2 t_3}_{r_1 r_2 r_3} = \int d\Omega_{\hat{\mathbf n}}
{Y}_{t_1 r_1} (\hat{\mathbf n}) {Y}_{t_2 r_2} (\hat{\mathbf n})
{Y}_{t_3 r_3} (\hat{\mathbf n}) \pkt
\ee
The indices $a,~b,~c$ correspond to the vector components and
repeated ones reflect the scalar product of the corresponding
vectors.

Similarly for the small scale approximation ($L<L_S$) we obtain
\bea \label{bispectrum4}
B^{m_1 m_2 m_3}_{l_1 l_2 l_3} &=&
\frac{i^{l_1+l_2+l_3-3} \sqrt{l_1(l_1+1) l_2(l_2+1) l_3(l_3+1)}}
{(2\pi^3)^3 [(L_\gamma/5) (\rho_{\gamma_0} + p_{\gamma_0}) ]^3 {\eta_0}^3}
 \int dk_1 dk_2 dk_3
j_{l_1}(k_1 \eta_0) j_{l_2}(k_2 \eta_0) j_{l_3}(k_3 \eta_0) \nn \\
&&\int d\Omega_{\hat{\mathbf k}_1} d\Omega_{\hat{\mathbf k}_2} d\Omega_{\hat{\mathbf k}_3}
\int d\Omega_{\hat{\mathbf q}} \int q^2 dq
M(|{\mathbf q}|) M(|{\mathbf k}_1-{\mathbf q}|) M(|{\mathbf k}_2+{\mathbf q}|) \nn \\
&&\sum_{t_1 r_1} \sum_{t_2 r_2} \sum_{t_3 r_3} i^{t_1+t_2+t_3} \int x^2 dx
j_{t_1}(k_1 x) j_{t_2}(k_2 x)  j_{t_3}(k_3 x)
 \ {G^{t_1 t_2 t_3}_{r_1 r_2 r_3}} \nn \\
&&\left[ {\mathbf Y}^{(-1)}_{t_1 r_1} (\hat{\mathbf k}_1) \times {\hat{\mathbf q}}\right]_{a}
\left[{{\mathbf Y}^{(1)}_{l_1 m_1}}^\star (\hat{\mathbf k}_1)\times(\hat{{\mathbf k}_1-\mathbf q})\right]_b \nn \\
&&\left[{\mathbf Y}^{(-1)}_{t_2 r_2} (\hat{\mathbf k}_2) \times \hat{\mathbf q}\right]_a
\left[{{\mathbf Y}^{(1)}_{l_2 m_2}}^\star (\hat{\mathbf k}_2) \times (\hat{{\mathbf k}_2+\mathbf q})\right]_c \nn \\
&&\left[{\mathbf Y}^{(-1)}_{t_3 r_3} (\hat{\mathbf k}_3) \times (\hat{{\mathbf k}_1 -\mathbf q})\right]_b
\left[{{\mathbf Y}^{(1)}_{l_3 m_3}}^\star (\hat{\mathbf k}_3) \times (\hat{{\mathbf k}_2+\mathbf q})\right]_c
\pkt
\eea

Using the Wigner $D$ functions and proceeding in the way analogous
to Sec. IIIA it can be shown that there are non-zero cross
correlations between non-equal $l$ and $m$. Again the proper answer
assumes accounting for the angular dependence in the power spectra
$M$.

\section{Concluding remarks}

We present the study of the magnetized perturbation vector mode
induced CMB two- and three-point correlation functions in a common
framework. We show that already the vorticity two-point correlation
functions reflect the anisotropy of the considered perturbations
before ensemble averaging. One of the implications of our
results is that CMB bispectrum computation technique presented in
Ref. \cite{Komatsu:2001rj} in the case of magnetized perturbations
should be applied with caution.

Note that in the present work we have focused on derivation of main
analytic results for the CMB two- and three-point correlation
functions arising from the vector mode supported by the stochastic
cosmological magnetic field. We plan to present a detailed analysis
of obtained equations and phenomenological estimates in a separate
publication.

\acknowledgments We are grateful to A.Kosowsky for fruitful
discussions on different points of present investigation (in
particular in deriving the ensemble averaging of the two-point
correlation function). We also appreciate useful communications and
comments from I. Brown, R. Durrer, and A. Lewis, and acknowledge
discussions with M. Kamionkowski,  B. Ratra, L. Samushia,  K.
Subramanian, and A. Tevzadze. We acknowledge partial support from Swiss National Science
Foundation SCOPES grant no. 128040, NSF grant AST1109180 and NASA Astrophysics Theory
Program grant NNXlOAC85G. T. K. 
acknowledges the ICTP associate membership program. 

\begin{appendix}

\section{Useful Mathematical Formulae}

In this Appendix we list various mathematical results we use in the computations.

\subsection{Wigner $D$ Functions}

Wigner $D$ functions relate helicity
basis vectors $ \mathbf{e'}_{\!\pm 1}=\mp (\mathbf{e}_\Theta
 \pm i\mathbf{e}_\phi)/\sqrt{2}$ and $\mathbf{e'}_{\! 0}=\mathbf{e}_r$
to  spherical basis vectors $\mathbf{e}_{\pm 1}=\mp (\mathbf{e}_x
\pm i\mathbf{e}_y)/\sqrt{2}$ and $\mathbf{e}_0=\mathbf{e}_z$ (see
Eq.~(53), p.~11,  \cite{varshalovich89}) through
\begin{eqnarray} \label{basis4}
\mathbf{e'}_{\!\!\mu}= \sum_\nu D^1_{\nu \mu} (\phi, \Theta, 0)
\mathbf{e}_\nu, ~~~~~\nu, \mu=-1,0,1 \pkt
\end{eqnarray}
In both  the spherical basis and the helicity basis the following
relations hold: $\mathbf{e}_\nu \mathbf{e}^\mu=\delta_{\nu \mu}$,
$\mathbf{e}^\mu =(-1)^\mu \mathbf{e}_{-\mu}$,
$\mathbf{e}^\mu=\mathbf{e}^\star_\mu$, $\mathbf{e}_\mu \times
\mathbf{e}_{\nu}=-i\epsilon_{\mu \nu \lambda} \mathbf{e}_\lambda$.

\subsection{Calculation of ${\cal B}$}\label{app_B}

For calculation of the two-point correlation function we must
determine the magnetic field energy momentum two-point cross
correlation $\langle \tau_{ab}^{(B)}({\bf k})
\tau_{cd}^{(B)}({\bf k^\prime })\rangle$
given by Eq. A6 of \cite{mkk02}
and Eq. 4.1 of \cite{bc05}. It is easy to show that the parts of
the $\tau^{(B)}_{ab}$ (and $\tau^{(B)}_{cd}$ proportional to the
$\delta_{cd}$), see Eq. (\ref{tau}),  do not contribute to the
integral of Eq. (\ref{pi}). Then we need to compute the following
object
\be
{\cal B}_{abcd} (\mathbf{k_1}, \mathbf{k_2})= \int
\frac{d^3 {\mathbf q_1}}{(2\pi)^3} \frac{d^3 {\mathbf
q_2}}{(2\pi)^3} \langle B_a^\star ({\mathbf q_1}) B_b^\star
({\mathbf k_1}-{\mathbf q_1}) B_c ({\mathbf q_2}) B_d ({\mathbf
k_2}-{\mathbf q_2}) \rangle \pkt \label{tau2}
\ee
Using Eq.~(\ref{spectrum}) and Wick's theorem, we obtain that the
contribution of the two-point correlation of the magnetic field
energy momentum into the vorticity perturbation is given by
\begin{eqnarray} \label{pi5}
\frac{ \delta({\bf k- k^\prime})}{(4\pi)^2}  \int d^3 q M(|{\bf q}|) M(|{\bf k - q}|)
\left[ P_{ac}({\bf \hat q}) P_{bd}({\bf {\hat{k-q}}}) +
P_{ad}({\bf \hat q}) P_{bc}({\bf {\hat{k-q}}}) \right] \pkt
\end{eqnarray}

For calculation of bispectrum we need to know following object
\be
{\cal B}_{abcdef} (\mathbf{k_1}, \mathbf{k_2}, \mathbf{k_3})= \int
\frac{d^3 {\mathbf q_1}}{(2\pi)^3} \frac{d^3 {\mathbf q_2}}{(2\pi)^3} \frac{d^3 {\mathbf q_3}}{(2\pi)^3}
\langle
B_a ({\mathbf q_1}) B_b ({\mathbf k_1}-{\mathbf q_1})
B_c ({\mathbf q_2}) B_d ({\mathbf k_2}-{\mathbf q_2})
B_e ({\mathbf q_3}) B_f ({\mathbf k_3}-{\mathbf q_3})
\rangle \pkt
\ee
Assuming that the magnetic field obeys Gaussian statistics we can expand the six-point correlation
function using Wick's theorem. Doing this we will get seven terms proportional to either
$\delta (\mathbf{k_1})$, $\delta (\mathbf{k_2})$ or $\delta (\mathbf{k_3})$ (which we neglect) and
eight terms proportional to $\delta (\mathbf{k_1}+\mathbf{k_2} +\mathbf{k_3})$ which we keep.
So, for ${\cal B}_{abcdef}$ one finds \cite{bc05,brown}:
\bea
{\cal B}_{abcdef} (\mathbf{k_1}, \mathbf{k_2}, \mathbf{k_3})&=&
\delta (\mathbf{k_1}+\mathbf{k_2} +\mathbf{k_3}) \int d^3 \mathbf{q}
M(|\mathbf{q}|) M(|\mathbf{k}_1-\mathbf{q}|) M(|\mathbf{k}_2+\mathbf{q}|)  \nn \\
&&\{
P_{ac} (\mathbf{q})
P_{be} (\mathbf{k}_1-\mathbf{q})
P_{df} (\mathbf{k}_2+\mathbf{q})
+
P_{ac} (\mathbf{q})
P_{bf} (\mathbf{k}_1-\mathbf{q})
P_{de} (\mathbf{k}_2+\mathbf{q})                                            \nn \\
&+&
P_{ad} (\mathbf{q})
P_{be} (\mathbf{k}_1-\mathbf{q})
P_{ef} (\mathbf{k}_2+\mathbf{q})
+
P_{ad} (\mathbf{q})
P_{bf} (\mathbf{k}_1-\mathbf{q})
P_{ce} (\mathbf{k}_2+\mathbf{q}) \}                                         \nn \\
&+&\delta (\mathbf{k_1}+\mathbf{k_2} +\mathbf{k_3}) \int d^3 \mathbf{q}
M(|\mathbf{q}|) M(|\mathbf{k}_1-\mathbf{q}|) M(|\mathbf{k}_3+\mathbf{q}|)   \nn \\
&&\{
P_{ae} (\mathbf{q})
P_{bc} (\mathbf{k}_1-\mathbf{q})
P_{df} (\mathbf{k}_3+\mathbf{q})
+
P_{ae} (\mathbf{q})
P_{bd} (\mathbf{k}_1-\mathbf{q})
P_{df} (\mathbf{k}_3+\mathbf{q})                                           \nn \\
&+&
P_{af} (\mathbf{q})
P_{bc} (\mathbf{k}_1-\mathbf{q})
P_{de} (\mathbf{k}_3+\mathbf{q})
+
P_{af} (\mathbf{q})
P_{bd} (\mathbf{k}_1-\mathbf{q})
P_{ce} (\mathbf{k}_3+\mathbf{q})
\} \pkt
\eea
Assuming that projector ${\cal A}^{abcdef}$ acting on this object is symmetric
w.r.t. each pair of indexes, i.e.
\be
{\cal A}^{abcdef}= {\cal A}^{bacdef}= {\cal A}^{abdcef}= {\cal A}^{abcdfe} \kma
\ee
quantity ${\cal B}_{abcdef}$ can be brought to a compact form
\be
{\cal B}_{abcdef} (\mathbf{k_1}, \mathbf{k_2}, \mathbf{k_3})=
8 \delta (\mathbf{k_1}+\mathbf{k_2} +\mathbf{k_3}) \int d^3 \mathbf{q}
M(|\mathbf{q}|) M(|\mathbf{k}_1-\mathbf{q}|) M(|\mathbf{k}_2+\mathbf{q}|)
P_{ac} (\mathbf{q})
P_{be} (\mathbf{k}_1-\mathbf{q})
P_{df} (\mathbf{k}_2+\mathbf{q}) \pkt
\ee

\subsection{Useful relations}

Following relation from vector algebra is valid
(see e.g. Eq.~(31), p.16 \cite{varshalovich89}):
\be \label{abcd}
[{\mathbf A} \times {\mathbf B}] \cdot [{\mathbf C} \times {\mathbf D} ] =
({\mathbf A} \cdot {\mathbf C}) \ ({\mathbf B} \cdot {\mathbf D}) -
({\mathbf A} \cdot {\mathbf D}) \ ({\mathbf B} \cdot {\mathbf C})
\pkt
\ee

Using plane wave decomposition into spherical harmonics one obtains
following useful representation for the Dirac delta function
\bea \label{delta}
\delta (\mathbf{k_1}+\mathbf{k_2}+\mathbf{k_3}) = \int \frac{d^3 x}{(2\pi)^3}
\e^{i(\mathbf{k_1}+\mathbf{k_2}+\mathbf{k_3}) {\mathbf x}} =
8 \int d^3 x
\sum_{t_1 r_1} \sum_{t_2 r_2} \sum_{t_3 r_3} i^{t_1+t_2+t_3}
j_{t_1}(k_1 x) j_{t_2}(k_2 x)  j_{t_3}(k_3 x) \nn \\
 { Y}_{t_1 r_1} (\hat{\mathbf k}_1)
{{ Y}^\star_{t_1 r_1}} (\hat{\mathbf x})
 { Y}_{t_2 r_2} (\hat{\mathbf k}_2)
{{ Y}^\star_{t_2 r_2}} (\hat{\mathbf x})
 { Y}_{t_3 r_3} (\hat{\mathbf k}_3)
{{ Y}^\star _{t_3 r_3}}(\hat{\mathbf x}) \pkt
\eea

\end{appendix}



\begin{thebibliography}{}
\bibitem{review}
N.~Bartolo, S.~Matarrese and A.~Riotto,
``Non-Gaussianity and the Cosmic Microwave Background Anisotropies,''
Adv.\ Astron.\  {\bf 2010}, 157079 (2010)
[arXiv:1001.3957 [astro-ph.CO]].

\bibitem{mukhanov92}
V.~F.~Mukhanov, H.~A.~Feldman and R.~H.~Brandenberger,
``Theory of cosmological perturbations. Part 1. Classical perturbations.
Part 2. Quantum theory of perturbations. Part 3. Extensions,''
Phys.\ Rept.\  {\bf 215}, 203 (1992).

\bibitem{WMAP7}
E.~Komatsu {\it et al.}  [WMAP Collaboration],
``Seven-Year Wilkinson Microwave Anisotropy Probe (WMAP) Observations:
Cosmological Interpretation,''
Astrophys.\ J.\ Suppl.\  {\bf 192}, 18 (2011)
[arXiv:1001.4538 [astro-ph.CO]].

\bibitem{Bernardeau:2002jf}
F.~Bernardeau and J.~P.~Uzan,
``Inflationary models inducing non-Gaussian metric fluctuations,''
Phys.\ Rev.\  D {\bf 67}, 121301 (2003)
[arXiv:astro-ph/0209330].

\bibitem{Komatsu:2010hc}
E.~Komatsu,
``Hunting for Primordial Non-Gaussianity in the Cosmic Microwave Background,''
Class.\ Quant.\ Grav.\  {\bf 27}, 124010 (2010).

\bibitem{Seshadri:2009sy}
T.~R.~Seshadri and K.~Subramanian,
``CMB bispectrum from primordial magnetic fields on large angular scales,''
Phys.\ Rev.\ Lett.\  {\bf 103}, 081303 (2009)
[arXiv:0902.4066 [astro-ph.CO]].

\bibitem{Caprini:2009vk}
C.~Caprini, F.~Finelli, D.~Paoletti and A.~Riotto,
``The cosmic microwave background temperature bispectrum from scalar
perturbations induced by primordial magnetic fields,''
JCAP {\bf 0906}, 021 (2009)

\bibitem{bc05}
I.~Brown and R.~Crittenden, ``Non-Gaussianity from Cosmic Magnetic Fields,''
Phys.\ Rev.\  D {\bf 72}, 063002 (2005) [arXiv:astro-ph/0506570];
I.~A.~Brown, ``Intrinsic Bispectra of Cosmic Magnetic Fields,''
Astrophys.\ J.\  {\bf 733}, 83 (2011) [arXiv:1012.2892 [astro-ph.CO]].

\bibitem{Trivedi:2010gi}
P.~Trivedi, K.~Subramanian and T.~R.~Seshadri,
``Primordial Magnetic Field Limits from Cosmic Microwave Background Bispectrum of
Magnetic Passive Scalar Modes,''
Phys.\ Rev.\ D {\bf 82}, 123006 (2010) [arXiv:1009.2724 [astro-ph.CO]].

\bibitem{subramanian98}
K.~Subramanian and J.~D.~Barrow,
``Magnetohydrodynamics in the early universe and the damping of noninear  Alfven waves,''
Phys.\ Rev.\  D {\bf 58}, 083502 (1998) [arXiv:astro-ph/9712083];
K.~Subramanian and J.~D.~Barrow,
``Microwave Background Signals from Tangled Magnetic Fields,''
Phys.\ Rev.\ Lett.\  {\bf 81}, 3575 (1998) [arXiv:astro-ph/9803261].

\bibitem{mkk02}
A.~Mack, T.~Kahniashvili and A.~Kosowsky,
``Microwave background signatures of a primordial stochastic magnetic field,''
Phys.\ Rev.\  D {\bf 65}, 123004 (2002).

\bibitem{lewis}
A.~Lewis,
``CMB anisotropies from primordial inhomogeneous magnetic fields,''
Phys.\ Rev.\  D {\bf 70}, 043011 (2004)

\bibitem{origin}
L.~M.~Widrow,
``Origin of Galactic and Extragalactic Magnetic Fields,''
Rev.\ Mod.\ Phys.\  {\bf 74}, 775 (2002);
A.~Kandus, K.~E.~Kunze and C.~G.~Tsagas,
``Primordial magnetogenesis,''
Phys.\ Rept.\  {\bf 505}, 1 (2011)  [arXiv:1007.3891 [astro-ph.CO]].

\bibitem{durrer}
R.~Durrer and C.~Caprini,
``Primordial Magnetic Fields and Causality,''
JCAP {\bf 0311}, 010 (2003).

\bibitem{ratra}
M.~S.~Turner and L.~M.~Widrow,
``Inflation Produced, Large Scale Magnetic Fields,''
Phys.\ Rev.\  D {\bf 37}, 2743 (1988);
B.~Ratra,
``Cosmological 'seed' magnetic field from inflation,''
Astrophys.\ J.\  {\bf 391}, L1 (1992);
K.~Bamba and M.~Sasaki,
``Large-scale magnetic fields in the inflationary universe,''
JCAP {\bf 0702}, 030 (2007).

\bibitem{limits}
D.~G.~Yamazaki, K.~Ichiki, T.~Kajino and G.~J.~Mathews,
``New Constraints on the Primordial Magnetic Field,''
Phys.\ Rev.\  D {\bf 81}, 023008 (2010);
D.~G.~Yamazaki, K.~Ichiki, T.~Kajino and G.~J.~Mathews,
``Constraints on the neutrino mass and the primordial magnetic field from the
matter density fluctuation parameter $\sigma_8$,''
  Phys.\ Rev.\  D {\bf 81}, 103519 (2010);
D.~Paoletti and F.~Finelli,
``CMB Constraints on a Stochastic Background of Primordial Magnetic Fields,''
Phys.\ Rev.\ D {\bf 83}, 123533 (2011) [arXiv:1005.0148 [astro-ph.CO]].

\bibitem{neronov}
A.~Neronov and I.~Vovk,
``Evidence for strong extragalactic magnetic fields from Fermi observations of TeV blazars,''
Science {\bf 328}, 73 (2010);
F.~Tavecchio, et al., F.~Tavecchio, G.~Ghisellini, L.~Foschini, G.~Bonnoli, G.~Ghirlanda and P.~Coppi,
`The intergalactic magnetic field constrained by Fermi/LAT observations of the TeV blazar 1ES 0229+200,''
Mon.\ Not.\ Roy.\ Astron.\ Soc.\ {\bf 406}, L70 (2010) arXiv:1004.1329 [astro-ph.CO];
F.~Tavecchio, G.~Ghisellini, G.~Bonnoli and L.~Foschini,
``Extreme TeV blazars and the intergalactic magnetic field,''
  arXiv:1009.1048 [astro-ph.HE];

A.~M.~Taylor, I.~Vovk and A.~Neronov,
``Extragalactic magnetic fields constraints from simultaneous GeV-TeV observations of blazars,''
Astron.\ Astrophys.\  {\bf 529}, A144 (2011) [arXiv:1101.0932 [astro-ph.HE]];
K.~Takahashi, M.~Mori, K.~Ichiki and S.~Inoue,
``Lower Bounds on Intergalactic Magnetic Fields from Simultaneously Observed GeV-TeV
Light Curves of the Blazar Mrk 501,''
arXiv:1103.3835 [astro-ph.CO];
C.~D.~Dermer, M.~Cavadini, S.~Razzaque, J.~D.~Finke and B.~Lott,
``Time Delay of Cascade Radiation for TeV Blazars and the Measurement of the
Intergalactic Magnetic Field,'' arXiv:1011.6660 [astro-ph.HE].

\bibitem{dolag} K.~Dolag, M.~Kachelriess, S.~Ostapchenko and R.~Tomas,
``Lower limit on the strength and filling factor of extragalactic magnetic fields,''
 Astrophys.\ J.\  {\bf 727}, L4 (2011)  [arXiv:1009.1782 [astro-ph.HE]];

\bibitem{Shiraishi:2010sm}
M.~Shiraishi, S.~Yokoyama, K.~Ichiki and K.~Takahashi,
``Analytic formulae of the CMB bispectra generated from non-Gaussianity in the tensor and vector perturbations,''
Phys.\ Rev.\ D {\bf 82}, 103505 (2010)
[arXiv:1003.2096 [astro-ph.CO]].

\bibitem{Shiraishi:2010yk}
M.~Shiraishi, D.~Nitta, S.~Yokoyama, K.~Ichiki and K.~Takahashi,
``Cosmic microwave background bispectrum of vector modes induced from primordial magnetic fields,''
Phys.\ Rev.\ D {\bf 82}, 121302 (2010)  [Erratum-ibid.\ D {\bf 83}, 029901 (2011)]
[arXiv:1009.3632 [astro-ph.CO]].

\bibitem{Shiraishi:2011fi}
M.~Shiraishi, D.~Nitta, S.~Yokoyama, K.~Ichiki and K.~Takahashi,
``Computation approach for CMB bispectrum from primordial magnetic fields,''
Phys.\ Rev.\ D {\bf 83}, 123523 (2011)
[arXiv:1101.5287 [astro-ph.CO]].

\bibitem{Shiraishi:2011dh}
M.~Shiraishi, D.~Nitta, S.~Yokoyama, K.~Ichiki and K.~Takahashi,
``Cosmic microwave background bispectrum of tensor passive modes induced from primordial magnetic fields,''
Phys.\ Rev.\ D {\bf 83}, 123003 (2011) [arXiv:1103.4103 [astro-ph.CO]].

\bibitem{bardeen}
J.~M.~Bardeen,
``Gauge Invariant Cosmological Perturbations,''
Phys.\ Rev.\  D {\bf 22}, 1882 (1980);
R.~Durrer,
``Gauge Invariant Cosmological Perturbation Theory: A General Study And Its
Application To The Texture Scenario Of Structure Formation,''
Fund.\ Cosmic Phys.\  {\bf 15}, 209 (1994). 

\bibitem{hw97}
W.~Hu and M.~J.~White,
``CMB Anisotropies: Total Angular Momentum Method,''
Phys.\ Rev.\  D {\bf 56}, 596 (1997).
[arXiv:astro-ph/9702170].

\bibitem{axel}
A.~Brandenburg, K.~Enqvist and P.~Olesen,
``Large-scale magnetic fields from hydromagnetic turbulence in the very early universe,''
Phys.\ Rev.\  D {\bf 54}, 1291 (1996) [arXiv:astro-ph/9602031].

\bibitem{jedamzik}
K.~Jedamzik, V.~Katalinic and A.~V.~Olinto,
``Damping of cosmic magnetic fields,''
Phys.\ Rev.\ D {\bf 57}, 3264 (1998)  [astro-ph/9606080].

\bibitem{dky98}
R.~Durrer, T.~Kahniashvili and A.~Yates,
``Microwave Background Anisotropies from Alfven waves,''
Phys.\ Rev.\  D {\bf 58}, 123004 (1998) [arXiv:astro-ph/9807089].

\bibitem{klr08}
T.~Kahniashvili, G.~Lavrelashvili and B.~Ratra,
``CMB Temperature Anisotropy from Broken Spatial Isotropy due to an
Homogeneous Cosmological Magnetic Field,''
Phys.\ Rev.\  D {\bf 78}, 063012 (2008).

\bibitem{varshalovich89}
D.~A.~Varshalovich, A.~N.~Moskalev, and V.~K.~Khersonskii,
{\it Quantum Theory of Angular Momentum}. (World Scientific, Singapore, 1988).

\bibitem{Komatsu:2001rj}
E.~Komatsu and D.~N.~Spergel,
``Acoustic signatures in the primary microwave background bispectrum,''
Phys.\ Rev.\  D {\bf 63}, 063002 (2001).

\bibitem{brown}
I.~Brown, 2009, private communication.

\end{thebibliography}
\end{document}